\documentclass[prl,nofootinbib,floatfix,twocolumn,showpacs,amsmath,amssymb,superscriptaddress]{revtex4}

\usepackage{graphicx}
\usepackage{dcolumn}
\usepackage{bm}
\usepackage{color}

\newcommand{\sparq}{{\ensuremath\stackrel{\rightharpoonup}{\Rightarrow}}}
\newcommand{\santq}{{\ensuremath\stackrel{\rightharpoonup}{\Leftarrow}}}
\newcommand{\de}{{\rm\,d}}
\newcommand{\detwo}{{\rm\,d^2\!}}

\def\hermes{{\sc Hermes}}

\def\ubar{\ensuremath{\overline{u}}}
\def\dbar{\ensuremath{\overline{d}}}
\def\sbar{\ensuremath{\overline{s}}}

\begin{document}

\title{
  Measurement of Parton Distributions of Strange Quarks in the Nucleon from
 Charged-Kaon Production in Deep-Inelastic Scattering on the Deuteron}


\def\groupargonne{\affiliation{Physics Division, Argonne National Laboratory, Argonne, Illinois 60439-4843, USA}}
\def\groupbari{\affiliation{Istituto Nazionale di Fisica Nucleare, Sezione di Bari, 70124 Bari, Italy}}
\def\groupbeijing{\affiliation{School of Physics, Peking University, Beijing 100871, China}}
\def\groupchina{\affiliation{Department of Modern Physics, University of Science and Technology of China, Hefei, Anhui 230026, China}}
\def\groupcolorado{\affiliation{Nuclear Physics Laboratory, University of Colorado, Boulder, Colorado 80309-0390, USA}}
\def\groupdesy{\affiliation{DESY, 22603 Hamburg, Germany}}
\def\groupzeuthen{\affiliation{DESY, 15738 Zeuthen, Germany}}
\def\groupdubna{\affiliation{Joint Institute for Nuclear Research, 141980 Dubna, Russia}}
\def\grouperlangen{\affiliation{Physikalisches Institut, Universit\"at Erlangen-
N\"urnberg, 91058 Erlangen, Germany}}
\def\groupferrara{\affiliation{Istituto Nazionale di Fisica Nucleare, Sezione di
 Ferrara and Dipartimento di Fisica, Universit\`a di Ferrara, 44100 Ferrara, Italy}}
\def\groupfrascati{\affiliation{Istituto Nazionale di Fisica Nucleare, Laborator
i Nazionali di Frascati, 00044 Frascati, Italy}}
\def\groupgent{\affiliation{Department of Subatomic and Radiation Physics, University of Gent, 9000 Gent, Belgium}}
\def\groupgiessen{\affiliation{Physikalisches Institut, Universit\"at Gie{\ss}en, 35392 Gie{\ss}en, Germany}}
\def\groupglasgow{\affiliation{Department of Physics and Astronomy, University of Glasgow, Glasgow G12 8QQ, United Kingdom}}
\def\groupillinois{\affiliation{Department of Physics, University of Illinois, Urbana, Illinois 61801-3080, USA}}
\def\groupmichigan{\affiliation{Randall Laboratory of Physics, University of Michigan, Ann Arbor, Michigan 48109-1040, USA }}
\def\groupmoscow{\affiliation{Lebedev Physical Institute, 117924 Moscow, Russia}
}
\def\groupnikhef{\affiliation{Nationaal Instituut voor Kernfysica en Hoge-Energiefysica (NIKHEF), 1009 DB Amsterdam, The Netherlands}}
\def\groupstpetersburg{\affiliation{Petersburg Nuclear Physics Institute, St. Petersburg, Gatchina, 188350 Russia}}
\def\groupprotvino{\affiliation{Institute for High Energy Physics, Protvino, Moscow region, 142281 Russia}}
\def\groupregensburg{\affiliation{Institut f\"ur Theoretische Physik, Universit\"at Regensburg, 93040 Regensburg, Germany}}
\def\grouprome{\affiliation{Istituto Nazionale di Fisica Nucleare, Sezione Roma
1, Gruppo Sanit\`a and Physics Laboratory, Istituto Superiore di Sanit\`a, 00161 Roma, Italy}}
\def\grouptriumf{\affiliation{TRIUMF, Vancouver, British Columbia V6T 2A3, Canada}}
\def\grouptokyo{\affiliation{Department of Physics, Tokyo Institute of Technology, Tokyo 152, Japan}}
\def\groupamsterdam{\affiliation{Department of Physics and Astronomy, Vrije Universiteit, 1081 HV Amsterdam, The Netherlands}}
\def\groupwarsaw{\affiliation{Andrzej Soltan Institute for Nuclear Studies, 00-689 Warsaw, Poland}}
\def\groupyerevan{\affiliation{Yerevan Physics Institute, 375036 Yerevan, Armenia}}
\def\groupnone{\noaffiliation}


\groupargonne
\groupbari
\groupbeijing
\groupchina
\groupcolorado
\groupdesy
\groupzeuthen
\groupdubna
\grouperlangen
\groupferrara
\groupfrascati
\groupgent
\groupgiessen
\groupglasgow
\groupillinois
\groupmichigan
\groupmoscow
\groupnikhef
\groupstpetersburg
\groupprotvino
\groupregensburg
\grouprome
\grouptriumf
\grouptokyo
\groupamsterdam
\groupwarsaw
\groupyerevan


\author{A.~Airapetian}  \groupmichigan
\author{N.~Akopov}  \groupyerevan
\author{Z.~Akopov}  \groupyerevan
\author{A.~Andrus}  \groupillinois
\author{E.C.~Aschenauer}  \groupzeuthen
\author{W.~Augustyniak}  \groupwarsaw
\author{R.~Avakian}  \groupyerevan
\author{A.~Avetissian}  \groupyerevan
\author{E.~Avetissian}  \groupfrascati
\author{S.~Belostotski}  \groupstpetersburg
% ...

\collaboration{The HERMES Collaboration} \noaffiliation

\date{\today} 

\begin{abstract}
The momentum and helicity density distributions 
of the strange quark sea 
in the nucleon are obtained in leading order from  
charged-kaon production in deep-inelastic scattering on the deuteron.
The distributions are extracted from spin-averaged
$K^{\pm}$ multiplicities, and from $K^{\pm}$  
and inclusive double-spin asymmetries for  
scattering of polarized positrons by a polarized
deuterium target. The shape of the momentum distribution is 
softer than that of the average of the $\ubar$ and $\dbar$ quarks. 
In the region of measurement $0.02<x<0.6$ and $Q^2>$1.0 GeV$^2$, the helicity
distribution is zero within experimental uncertainties. 
\end{abstract}



\maketitle

Parton distribution functions (PDFs) form the basis for the description
of the flavor structure of the nucleon. 
The spin-averaged parton distribution 
functions $q(x)$ of quarks and antiquarks
of flavors $q=(u,d,s)$ \cite{pdf:cteq5,pdf:MRST2001} describe
the quark momentum contributions, where $x$ is the dimensionless
Bjorken scaling variable representing the momentum fraction of
the target carried by the parton in a frame where the target has 
``infinite" longitudinal momentum.
They are sums of the number densities of the quarks
$q_\sparq(x)$ [$q_\santq(x)$]
with the same [opposite] helicity as that of the nucleon. 
The differences, or helicity distributions,
$\Delta q(x)=q_\sparq(x)-q_\santq(x)$ 
describe the flavor dependent contributions of the quark spins 
to the spin of the nucleon. 
The features of the parton distributions
reflect the QCD dynamics of the constituents.
Because strange quarks are objects which reflect directly properties of
the nucleon sea, they are of special interest. 
Their distributions are
also important because of their impact on quantitative calculations of
certain key short-distance processes at hadron colliders, and their
implications for the measurement of the Weinberg angle in deep-inelastic 
scattering (DIS) of neutrinos~\cite{pdf:cteq6a,nutev:thetaw}.

In the absence of significant experimental constraints, current global
QCD fits of PDFs~\cite{pdf:kretzer,pdf:thorne} assume 
the strange quark and antiquark momentum
distributions $s(x)$ and $\sbar (x)$ to be given by 
$s(x)=\sbar (x)=r[\ubar (x)+\dbar (x)]/2$ with $r\approx 1/2$ at some
low factorization scale.
Measurements of neutrino and antineutrino production of 
dimuons~\cite{nu:cdhs,nu:ccfr1,nu:ccfr2,nu:charm,nu:nomad,nutev:tzanov,
nu:bebc,nu:e531,nu:chorus}
provide useful but 
limited information~\cite{pdf:lai} on the 
normalization and shape of the distribution 
$s(x)+\sbar (x)$. In these experiments, extraction of the strange quark
distributions requires knowledge of the charm quark mass, the charm hadron
semileptonic branching ratio, and the ``Peterson fragmentation parameter''
~\cite{nu:peter} that describes the kinematic dependence of the
charm fragmentation function. These quantities together with the strange 
parton distributions themselves are fitted simultaneously in the extraction
procedure.  Much of the information on 
properties of the helicity
distribution of strange quarks is based on the analysis of inclusive 
DIS and hyperon decay under the assumption 
of SU(3) symmetry among the structures of the octet baryons. 
In these inclusive experiments~\cite{emc:g1} the first moment of the
helicity distribution for strange quarks is one of the principal results.
The most precise recent value is $-0.103\pm0.007(exp.)\pm 0.013(theor.)\pm
0.008(evol.)$ in LO~\cite{hermes:g1dlong}. 
A full 5-flavor decomposition using \hermes\ semi-inclusive 
DIS~\cite{hermes:deltaq} data from proton and deuteron
targets, although not sensitive to 
$\Delta\sbar (x)$, yielded $\Delta s=0.028\pm 0.033\pm 0.009$
for the first partial moment of the strange quark helicity density
in the measured range $0.023<x<0.3$. A separate ``isoscalar''
extraction of $\Delta s+\Delta \sbar$ from DIS data on the deuteron alone
gave $\Delta s+\Delta\sbar = 0.129\pm 0.042\pm 0.129$ 
in the measured range where the large
systematic uncertainty reflected lack of knowledge of kaon 
fragmentation functions.

This letter reports a new isoscalar extraction of 
$s(x)+\sbar (x)$ and $\Delta (s(x)+\sbar (x))$ based on the same
\hermes\ data obtained from polarized DIS on a deuterium target.
The measurement reported here is complementary to the neutrino
results, and is the first extraction of $s(x)+\sbar (x)$ in charged lepton
DIS.
Because strange quarks carry no isospin, the strange seas in the proton and
neutron can be assumed to be identical. 
In the deuteron, an isoscalar target, the fragmentation
process in DIS can be described by fragmentation functions that have
no isospin dependence. Aside from isospin
symmetry between proton and neutron, the only symmetry assumed is 
charge-conjugation invariance in fragmentation.
For the isoscalar deuteron in Leading Order (LO), 
the inclusive unpolarized (U) electron 
scattering cross section in terms of the parton distributions
$Q(x)\equiv u(x)+\ubar (x)+d(x)+\dbar (x)$ and $S(x)\equiv s(x)+\sbar (x)$
takes the form
\begin{eqnarray}
\label{eq:sigUU}
\frac{\detwo N^{DIS}(x)}{\de x \de Q^2}= 
{\cal K}_{U}(x,Q^2) \left[5 Q(x) + 2 S(x) \right] \,,
\end{eqnarray}
where ${\cal K}_{U}(x,Q^2)$ is a kinematic factor containing the 
hard scattering cross section.
\noindent
The weak logarithmic dependence of the PDFs 
on $-Q^2$, the squared four-momentum
of the exchanged virtual photon, has been suppressed for simplicity.
Applying the same LO formalism to the semi-inclusive cross section for
charged kaon production, irrespective of charge, hereafter 
designated as $K$ gives
\begin{eqnarray}
\label{eq:KsigUU}
\lefteqn{\frac{\detwo N^{K}(x)}{\de x \de Q^2} =
{\cal K}_{U}(x,Q^2)\times}   \nonumber  \\ 
 & &  \left[Q(x) {\int{\cal D}^{K}_{Q}(z)dz} 
+ S(x){\int{\cal D}^{K}_{S}(z)dz} \right] \,,
\end{eqnarray}
\noindent
where $z\equiv E_h/\nu$ with $\nu$ and $E_h$ the energies  of the virtual 
photon and of the detected hadron in the target rest frame,
 ${\cal D}^K_{Q}(z)  \equiv  4{D}_u^{K}(z)+
{D}_d^{K}(z)$ and
${\cal D}^K_{S}(z) \equiv  2{D}_s^{K}(z).$
The fragmentation function $D_q^K(z)$ describing 
the number density of charged kaons
from a struck quark of flavor $q$ is integrated over the measured range of $z$.
Combining Eqs. (\ref{eq:sigUU},\ref{eq:KsigUU}) 
and neglecting the term $2 S(x)$ compared to $5 Q(x)$,
it follows immediately that
\begin{equation}\label{eq:scomp}
 S(x){\int{\cal D}^{K}_{S}(z)dz}\simeq Q(x)\left[5
\frac{\detwo N^{K}(x)}{\detwo N^{DIS}(x)}-
{\int{\cal D}^{K}_{Q}(z)dz}\right]\, .
\end{equation} 
Eq.~\ref{eq:scomp} is the basis for the extraction of the
quantity $S(x)\int D_S^{K}(z)dz$.

The data were recorded with a longitudinally nuclear-polarized deuteron 
gas target internal to the $E$ = 27.6 GeV {\sc Hera} positron storage ring at
{\sc Desy}. The self-induced beam polarization was
measured continuously with Compton backscattering of circularly polarized
laser beams~\cite{hermes:tpol2,hermes:lpol}. 
The open-ended target cell was fed by an
atomic-beam source based on Stern-Gerlach separation with
hyperfine transitions. 
The nuclear polarization of the atoms was flipped at 90\,s time intervals,
while both this polarization and the atomic fraction inside the target cell 
were continuously measured~\cite{hermes:TG}.  
The average value of the deuteron polarization 
was 0.845 with a fractional systematic uncertainty of 3.5\%.

Scattered beam leptons and coincident hadrons were detected by
the \hermes\ spectrometer~\cite{hermes:spectr}. Leptons were identified
with an efficiency exceeding 98\% and a hadron
contamination of less than 1\% using 
an electromagnetic calorimeter, a transition-radiation detector, a 
preshower scintillation counter and a ring-imaging 
{\v C}erenkov (RICH) detector~\cite{hermes:rich}. The dual-radiator
RICH was also used to identify charged kaons. 
Events were selected subject to the kinematic
requirements $Q^2 >~ 1$\,GeV$^2$, $W^2 > 10$\,GeV$^2$ and $y < 0.85$,
where 
$W$ is the invariant mass of the 
photon-nucleon system, and $y=\nu/E$.
Coincident hadrons were accepted if $0.2<z<0.8$
and $x_F\approx 2p_L/W>0.1$, where $p_L$ is the longitudinal
momentum of the hadron with respect to the virtual photon
direction in the photon-nucleon center of mass frame.
The Bjorken $x$ range of measurement was 0.02--0.6.

\begin{figure}[h]
\includegraphics[width=8cm]{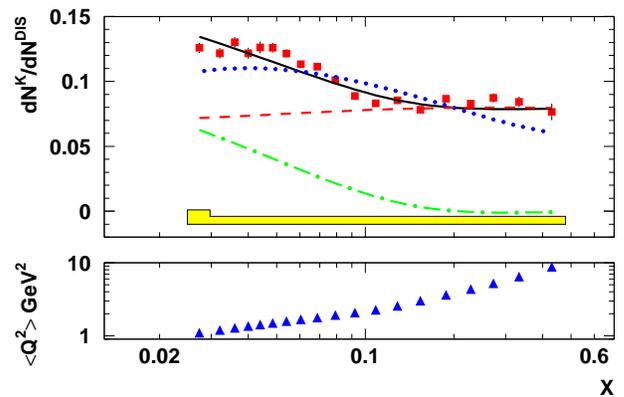}
\caption{\label{fig:mult_kaons} The multiplicity 
corrected to 4$\pi$ of charged kaons
in semi-inclusive DIS from a  deuterium target, as a
function of Bjorken $x$. The continuous curve 
is calculated from the curve in Fig.~\ref{fig:sxdz} using
Eq. \ref{eq:scomp}.
The dashed(dash-dotted) curve is the nonstrange(strange) quark contribution 
to the multiplicity for this fit.
The dotted curve is the best fit to $\int{\cal D}^{K}_{S}(z)dz$
using {\sc Cteq6l} PDFs.
The error bars are statistical. The band represents the systematic 
uncertainties.
The values of $\langle Q^2 \rangle$ for
each $x$ bin are shown in the lower panel.}
\end{figure}

\begin{figure}[!b]
\centerline{\includegraphics[width=8cm]{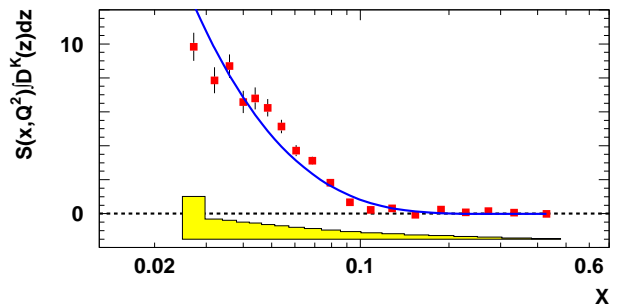}}
\caption{\label{fig:sxdz} The strange fragmentation product
$S(x,Q^2){\int{\cal D}^{K}_{S}(z)dz}$ obtained from the measured
\hermes\ multiplicity for charged kaons at the $\langle Q^2 \rangle$
for each bin. The curve is a least squares
fit of the form $x^{-0.863}e^{-x/0.0487}(1-x)$. 
The band represents systematic uncertainties.}
\end{figure}

The charged kaon multiplicity was extracted by summing over the
kaon yields for the two beam-target polarization states. 
An event weighting procedure was used to correct for RICH kaon 
identification inefficiencies.  
The effects of QED radiation, instrumental resolution,
and acceptance were 
simulated~\cite{ph:pepsi,hermes:radgen,ph:jetset}, 
and corrections were applied 
to the data for each polarization state using
a technique that unfolds kinematic migration of events~\cite{hermes:g1dlong}.
The results 
are presented in Fig.~\ref{fig:mult_kaons}.   
The trends in the data were not reproduced (see dotted curve in
Fig.~\ref{fig:mult_kaons}) by fitting the points using 
the {\sc Cteq6l}~\cite{pdf:cteq6} strange quark PDFs
in Eqs. \ref{eq:sigUU} and \ref{eq:KsigUU}, with
$\int{\cal D}^{K}_{Q}(z)dz$ and $\int{\cal D}^{K}_{S}(z)dz$
as free parameters.
In view of the paucity of reliable data on $S(x)$, it was
assumed instead that it is unknown, 
and the analysis was carried out extracting
the product $S(x)\int D_S^{K}(z)dz$ in LO. 
For $x>0.15$ the multiplicity is constant at a value of about 0.080,
implying that $S(x)/Q(x)$ is constant. For this analysis
$S(x)$ is assumed to be negligible at large $x$ from which
it follows that $S(x)=0$ 
for $x>0.15$ and that
$\int_{0.2}^{0.8}{\cal D}^K_Q(z)dz=0.398\pm 0.010$, in excellent 
agreement with the value $0.435\pm 0.044$
 obtained for $Q^2=2.5$ GeV$^2$ from the most
recent global analysis of fragmentation functions~\cite{ff:deflorian}.
The value 0.398 was then used in Eq. (\ref{eq:scomp}) together
with values of $Q(x)$ from {\sc Cteq6l} and the measured multiplicities 
to obtain the product $S(x){\int{\cal D}^{K}_{S}(z)dz}$
shown in Fig.~\ref{fig:sxdz}. 
A small iterative correction was made to account for the neglect
of the $2S(x)$ term in Eq.~\ref{eq:sigUU}.
The result for the product together with a fit 
of the form $x^{-a_1}e^{-x/a_2}(1-x)$ is shown in Fig.~\ref{fig:sxdz}, 
and leads to the continuous curve in Fig.~\ref{fig:mult_kaons}.

\begin{figure}[t]\centerline{
\includegraphics[width=8cm]{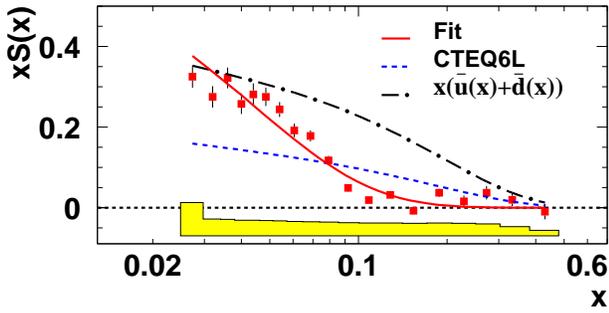}}
\caption{\label{fig:sxdz2p5} 
The strange parton distribution $xS(x)$ from
the measured \hermes\ multiplicity for charged kaons evolved to
$Q_0^2=2.5$ GeV$^2$ assuming ${\int{\cal D}^{K}_{S}(z)dz}=1.27\pm 0.13$. 
The solid curve is a 3-parameter
fit for $S(x)=x^{-0.924}e^{-x/0.0404}(1-x)$, the dashed curve
gives $xS(x)$ from {\sc Cteq6l}, and the dot-dash curve is the sum of light
antiquarks from {\sc Cteq6l}.}
\end{figure}

The improved fit (continuous curve in Fig.~\ref{fig:mult_kaons})
to the multiplicity 
is an indication
that the actual distribution of $S(x)$ is substantially different from the
average of those of the nonstrange antiquarks. To explore this point, 
the \hermes\ result for
$S(x){\int{\cal D}^{K}_{S}(z)dz}$ has been evolved to 
$Q^2_0 = 2.5$ GeV$^2$. The $Q^2$ evolution factors were taken 
from {\sc Cteq6l} and the fragmentation function compilation given 
in~\cite{ff:deflorian}. 
Consideration of corrections to the evolution due to higher twist 
contributions is not necessary, since higher twist effects are
expected to be significant~\cite{pdf:martin} only for larger values
of $x$ where the extracted distribution of $xS(x)$ vanishes.
The distribution of $xS(x)$ was 
obtained from $S(x){\int{\cal D}^{K}_{S}(z)dz}$ by dividing by
${\int{\cal D}^{K}_{S}(z)dz}=1.27\pm 0.13$, 
the value at $Q^2=2.5$ GeV$^2$ given  
in~\cite{ff:deflorian}.  
The results are presented in Fig.~\ref{fig:sxdz2p5}. 
The normalization of 
the \hermes\ points is determined by the value
of ${\int{\cal D}^{K}_{S}(z)dz}$ assumed. However, whatever the 
normalization, the shape of $xS(x)$ implied 
by the \hermes\ data is incompatible
with $xS(x)$ from {\sc Cteq6l} as well as 
the assumption of an average of an isoscalar nonstrange sea. 
The absence of strength above $x\approx 0.1$ is clearly discrepant with
{\sc Cteq6l}, while deviations from the {\sc Cteq6l} 
prediction at low $x$ could be, in part, a manifestation of higher order
processes. 

In the isoscalar extraction of the helicity distribution 
$\Delta S(x)=\Delta s(x)+\Delta\sbar(x)$, only the double-spin 
asymmetry $A_{\parallel ,d}^{K}(x,Q^2)$
for all charged kaons, irrespective of charge, and the inclusive 
asymmetry $A_{\parallel ,d}(x,Q^2)$ are used.  
In LO, the inclusive and the charged kaon 
double-spin(LL) asymmetries are determined by 
the relations
\begin{eqnarray}
\label{eq:sigLL}
\lefteqn{A_{\parallel ,d}(x) \frac{\detwo N^{DIS}(x)}{\de x \de Q^2} }
\nonumber \\
 & & = {\cal K}_{LL}(x,Q^2) \left[5 \Delta Q(x) + 
2 \Delta S(x)\right] \, ,
\end{eqnarray}
where ${\cal K}_{LL}$ is a kinematic factor, and
\begin{eqnarray} \label{eq:KsigLL}
\lefteqn{A^{K^{\pm}}_{\parallel ,d}(x) \frac{\detwo N^{K}(x)}
{\de x \de Q^2} = {\cal K}_{LL}(x,Q^2)\times}\nonumber \\
 & &  \left[\Delta Q(x) 
{\int{\cal D}^{K}_{Q}(z)dz} + \Delta S(x) {\int{\cal D}^{K}_{S}(z)dz}
\right] \,.
\end{eqnarray}
Eqs. (\ref{eq:sigLL},\ref{eq:KsigLL}) permit the simultaneous
extraction of the helicity distribution $\Delta Q(x)=\Delta u(x)+
\Delta \ubar (x)+\Delta d(x) + \Delta\dbar (x)$ and the strange
helicity distribution $\Delta S(x)=\Delta s(x)+\Delta\sbar (x)$.
The nonstrange integrated fragmentation function needed for a 
LO extraction of $\Delta 
S(x)$ was extracted from the multiplicity analysis of the same data.

\begin{figure}[t]
\includegraphics[width=8cm]{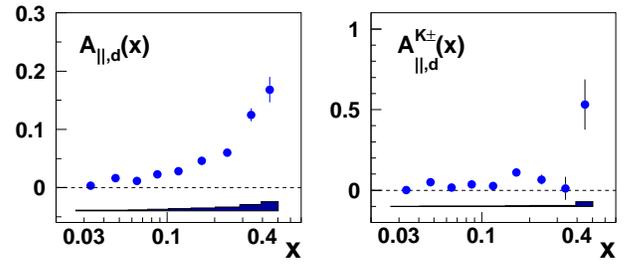}
\caption{\label{fig:A1d} Lepton-nucleon polarized cross section asymmetries
$A_{\parallel,d}$ for inclusive DIS and
$A^{K}_{\parallel,d}$ for semi-inclusive DIS by a deuteron target as a
function of Bjorken $x$, for identified charged kaons.
The error bars are statistical, and the bands at the
bottom represent the systematic uncertainties.}
\end{figure}

The semi-inclusive asymmetries $A^{K}_{\parallel,d}$ 
were derived from the kaon spectra
measured for each target polarization. The target polarization 
was corrected for 
the D-wave admixture in the deuteron wave function by applying
the correction term $(1-1.5\omega_D)$ 
where $\omega =0.05\pm0.01$~\cite{ph:D-state:machleidt}.
The corrected asymmetries are shown in Fig.~\ref{fig:A1d}. 
The inclusive asymmetries $A_{\parallel,d}(x)$ were corrected for effects
of QED radiation and instrumental smearing with the same 
procedures described above for the spin dependent kaon multiplicities. 
Contributions
to the systematic uncertainties in the asymmetries include those from the
beam and target polarizations, and  
the neglect of the transverse spin structure function
$g_2(x)\approx 0$~\cite{E155:g2}, and for 
$A^{K}_{\parallel,d}$ from those of RICH kaon identification.

\begin{figure}[!t]
\centerline{\includegraphics[width=8cm]{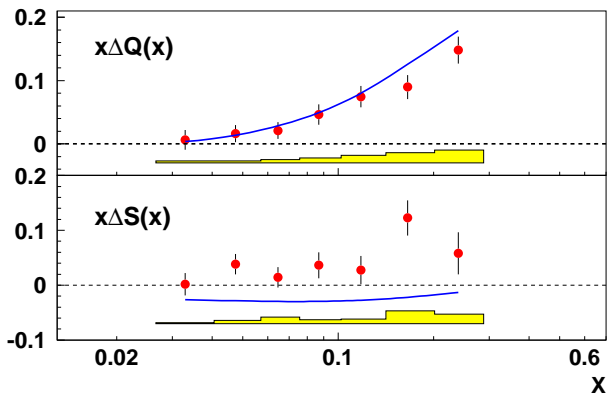}}
\caption{\label{fig:helicities2} Nonstrange and strange quark helicity
distributions at $Q^2_0=2.5$\,GeV$^2$, as a function of 
Bjorken $x$.
The error bars are statistical, and the bands at the
bottom represent the systematic uncertainties. The curves are the LO 
results of Leader {\emph{et al.}}\cite{pdf:leader} from their  
analysis of world data.}
\end{figure}

The quark helicity distributions 
were extracted from the measured 
spin asymmetries $A_{\parallel,d}(x)$ and $A^{K}_{\parallel,d}(x)$ 
in an analysis based on Eqs. (\ref{eq:sigLL},\ref{eq:KsigLL}).
The value of ${\int{\cal D}^{K}_{S}(z)dz}=1.27\pm 0.13$ was used to extract
$\Delta S(x)$. The results are presented in Fig.~\ref{fig:helicities2}.
The strange helicity distribution also
agrees well with the less precise results 
of~\cite{hermes:deltaq}, and is consistent with
zero over the measured range.

The first moments of the helicity densities in the measured region 
are presented in Tab.~\ref{tab:moments}. The result for 
$\Delta Q$ over the measured range is consistent with the value 
$0.381\pm 0.010({\rm stat.})\pm0.027({\rm sys.})$ 
for the full moment previously extracted from \hermes\ $g_{1,d}$
data~\cite{hermes:g1dlong}.
The value of $\Delta S$ measured here is not in serious disagreement with 
$-0.0435\pm 0.010({\rm stat.}) \pm 0.004({\rm sys.})$
 extracted from the inclusive {\hermes} measurements.
The value for the partial moment of the octet combination 
$\Delta q_8(x)=\Delta Q(x) -2\Delta S(x)$, 
included in Tab.~\ref{tab:moments}, is substantially less
than the value of the axial charge $a_8
\equiv \Delta q_8=\int_0^1\Delta q_8(x)dx=0.586 \pm 0.031$ 
extracted from  
the hyperon decay constants by assuming SU(3) symmetry~\cite{ph:ratcliffe}.
Possible explanations for the deficit observed for $\Delta q_8$ include
violation of SU(3) symmetry or missing octet strength at values of $x$ below
the measured range. The substantial deviation
observed in the shape of $S(x)$ from that of the light sea quarks is a clear
manifestation of violation of SU(3) symmetry
 \cite{ph:lichten-lipkin,ph:leader,ph:weigel} in the strange quark sector.

In conclusion, inclusive and semi-inclusive-charged-kaon spin asymmetries
for a longitudinally polarized deuteron target have been analyzed to
extract the LO parton distributions of the strange sea in the
proton. The partial moment of the nonstrange fragmentation function
needed for the LO analysis has been extracted directly from the same data.
The values for the PDFs presented in this paper are available
at the \hermes\ web site (http://www-hermes.desy.de).
The momentum densities are softer than previously assumed.
The helicity densities are consistent with zero and
the partial moment of the octet axial combination is observed to be
substantially less than the axial charge
extracted from hyperon decays under the
assumption of SU(3) symmetry.
\begin{table}[h]
  \renewcommand{\arraystretch}{1.1}
  \caption{\label{tab:moments} First moments of various helicity
    distributions in the Bjorken $x$ range 0.02--0.6 
    at a scale of
    \mbox{$Q_0^2 = 2.5\:\text{GeV}^2$.}}
      \begin{ruledtabular}
    \begin{tabular}{lr} 
      & \multicolumn{1}{c}{Moments in measured range} \\ \hline
      $\Delta Q$ \rule{0mm}{3ex} & $ 0.359 \pm 0.026({\rm stat.}) \pm 0.018({\rm sys.})$ \\
      $\Delta S$                 & $ 0.037 \pm 0.019({\rm stat.}) \pm 0.027({\rm sys.})$ \\
      $\Delta q_8$               & $ 0.285 \pm 0.046({\rm stat.}) \pm 0.057({\rm sys.})$ \\
    \end{tabular}
  \end{ruledtabular}
\end{table}

We gratefully acknowledge the {\sc Desy} management for its support, the staff
at {\sc Desy} and the collaborating institutions for their significant effort, 
and our national funding agencies for their financial support. We thank D. 
de Florian, R. Sassot, and M. Stratmann for discussions and the early use of
their programs for calculating PDFs and fragmentation functions. 

\bibliography{dsletter}

\end{document}